\documentclass{raa} 

\usepackage{epsfig ,graphicx}
\usepackage{txfonts}
\usepackage{natbib}
\bibpunct{(}{)}{;}{a}{}{,} 
\usepackage{natbib,twoopt}

\begin{document}

\title{Numerical Simulation of a possible origin of the positive radial metallicity gradient of the thick disk}

\author{Awat Rahimi\inst{1},
 Kenneth Carrell\inst{1}, 
 Daisuke Kawata\inst{2}
 }

\institute{National Astronomical Observatories, Chinese Academy of Sciences, Beijing 100012, China\\
\email{ara@nao.cas.cn}
\and
Mullard Space Science Laboratory, University College London, Holmbury St. Mary, Dorking, Surrey, RH5 6NT, UK
}

  \date{Received August 9, 2013; Accepted YYY, 2013}

\abstract{
We analyze the radial and vertical metallicity and [$\alpha$/Fe] gradients of the disk stars of a disk galaxy simulated in a fully cosmological setting with the chemodynamical galaxy evolution code, {\tt GCD+}. We study how the radial abundance gradients vary as a function of height above the plane and find that the metallicity ([$\alpha$/Fe]) gradient becomes more positive (negative) with increasing height, changing sign around 1.5 kpc above the plane. At the largest vertical height (2 $<$ $|z|$ $<$ 3 kpc), our simulated galaxy shows a positive radial metallicity gradient. We find that the positive metallicity gradient is caused by the age-metallicity and age-velocity dispersion relation, where the younger stars have higher metallicity and lower velocity dispersion. Due to the age-velocity dispersion relation, a greater fraction of younger stars reach $\left | z \right |>2$ kpc at the outer region, because of the lower gravitational restoring force of the disk, i.e. flaring. As a result, the fraction of younger stars with higher metallicity due to the age-metallicity relation becomes higher at the outer radii, which makes the median metallicity higher at the outer radii. Combining this result with the recently observed age-metallicity and age-velocity dispersion relation for the Milky Way thick disk stars suggested by \citet{hdl13}, we argue that the observed (small) positive radial metallicity gradient at large heights of the Milky Way disk stars can be explained by the flaring of the younger thick and/or thin disk stars.}


\titlerunning{
Metallicity gradients in the thick disk
}
\authorrunning{Rahimi et al.}
\maketitle

\section{Introduction}
\label{intro-sec}

The formation mechanism of the Milky Way (MW) and its individual components remain an important outstanding problem in modern Astrophysics.
\citet{gilr83} showed that near the Sun, the stellar vertical density profile is tightly matched by the sum of two exponentials in $|z|$ with scaleheights of 300 and 900 pc \citep{jur08}. This observation led to the notion of a disk which was made up of two distinct (but overlapping) components, the thin and thick disks. The vertical scaleheight of the thick disk is significantly larger than the thin disk, but its precise value is uncertain, making it difficult to determine the fraction of local stars which belong to each component. Current estimates put the value around 10:1 in favor of the thin disk \citep{jur08}. 

The authenticity of the distinction between the thin and thick disks is still debated \citep[e.g.][]{isj08,brd12}, and an important aspect of recent research has been to find a method of separating the two components in a physically well-motivated way.
A kinematical separation like that used by \citet{bfl05} has had some success as 
it is argued that thick disk stars can be identified by their lower rotational velocity, since near the Sun, the thick disk lags solar rotation.
Perhaps a clearer distinguishing criterion is chemistry. Thick disk stars generally have, at any given metallicity [Fe/H], a higher abundance of alpha$-$elements to iron ratio [$\alpha$/Fe] \citep[e.g.][]{fuh08}.
This high [$\alpha$/Fe] value implies that the thick disk formed rapidly within the 
 early stages of the Galaxy's life.
Thick disk stars therefore are thought to be very old. 
Studying the chemical properties of stars can provide important constraints and boundary conditions for models of Galaxy formation.
There are several theories for the origin of the thick disk including disk heating \citep{kroupa02}, satellite accretion \citep{anse03b,yd05}, high redshift gas rich mergers \citep{bkgf04b}, radial migration \citep{rdq08,sbin09b} and a combination of high redshift gas rich mergers and radial migration \citep{bsgk12,mcm12}. In all formation scenarios, the thick disk is older than the thin disk. 
Thick disks have been observed in many other disk galaxies \citep[e.g.][]{db02,yd06} suggesting that they may be ubiquitous.

\citet{lrd11} recently showed using $N$-body SPH simulations that many properties of the thick disk can be explained by the radial migration of old, $\alpha$-rich stars from the inner to the outer regions of the disk. Using a suite of self-consistent cosmological simulations, \citet{pfg12} examined the evolution of metallicity gradients finding most models predict radial gradients consistent with those observed in late-type disks at the present day.
Using an isolated $N$-body simulation of an evolving thick disk model, \citet{css14} analyzed the effects of dynamical evolution on the initial chemical conditions and found that the chemical imprint at the time of disk formation is not washed out by secular dynamical processes.


In this paper, we analyze for the first time both the metallicity and [$\alpha$/Fe] gradients in both the radial and vertical directions for a cosmologically simulated disk galaxy. 
The major benefit of this study (in addition to being cosmological) is the detailed chemical evolution model which allows us to qualitatively compare the radial and vertical [Fe/H] gradients in our disk galaxy with recent observations from the MW, and allows us to make predictions as to the expected [$\alpha$/Fe] gradients. We especially highlight the radial metallicity gradient at high vertical height, and provide a possible mechanism to cause a positive metallicity gradient. 

\section{The code and model}
\label{code-sec}

The code used to simulate this galaxy (GAL 1) has already been described in Rahimi et al. (2011). Here we very briefly summarize the main points and refer the reader to Rahimi et al. (2011) for the details\footnote{Also, for the reader interested in the analytical arguments, we refer them to the previous literature \citep[e.g.][]{cmr01,rkf05,mmc10}.}. 
We use the original galactic chemodynamical evolution code {\tt GCD+} developed by \citet{kg03a}. 
{\tt GCD+} is a three-dimensional tree $N$-body/smoothed particle hydrodynamics code   \citep{ll77,gm77,bh86,hk89,kwh96} that incorporates self-gravity, hydrodynamics, radiative cooling, star formation, supernova feedback, metal enrichment and metal-dependent radiative cooling \citep[derived with {\tt MAPPINGSIII}:][]{sd93}.
However, we ignored the effect of the cosmic UV background radiation and UV radiation from hot stars for simplicity.

For star formation to occur, the following criteria must be satisfied: (i) the gas density is greater than some threshold density, taken to be 0.1 $\rm cm^{-3}$ ; (ii) the gas velocity field is convergent and (iii) the  gas is locally Jeans unstable. 
Our star formation formula corresponds to the Schmidt law. We assume that stars are distributed according to the standard \citet{s55} initial mass function (IMF).
We assume pure thermal feedback from both Type II (SNe II) and Type Ia (SNe Ia) supernovae (SNe) with $10^{50}$ erg/SN being coupled to the surrounding gas particles.

{\tt GCD+} takes into account the metal-dependent nucleosynthetic byproducts of SNe II \citep{ww95}, SNe Ia \citep{ibn99}, and intermediate-mass AGB stars \citep{vdhg97}. 
We adopt the \citet{ktn00} SNe Ia progenitor formalism.   

\begin{table*}
 \begin{minipage}{140mm}
  \caption{Simulation parameters}
  \renewcommand{\footnoterule}{}
  \begin{tabular}{@{}lllllllllll@{}}
  \hline
   Name & $M_{\rm vir}$ & $r_{\rm vir}$ & $m_{\rm gas}$\footnote{Mass of gas per particle} & $m_{\rm DM}$ \footnote{Mass of dark matter per particle} & $e_{\rm gas}$ & $e_{\rm DM}$ & $\Omega_0$ & $h_0$ & $\Omega_{\rm b}$ \\
   & (M$_{\odot }$) & $(kpc)$ & (M$_{\odot }$) & (M$_{\odot }$) & {$(kpc)$} & {$(kpc)$} & & & & \\
   \hline
GAL 1 & $8.8\times10^{11}$ & 240 & $9.2\times10^{5}$ & $6.2\times10^{6}$ & 0.57 & 1.1 & 0.3 & 0.7 & 0.039  \\
 \hline
\end{tabular}
\end{minipage}
\end{table*}

The galaxy simulated here is from the sample of \citet{rah09}, referred to as ``GAL 1". GAL 1 is a high-resolution version of galaxy ``D1" in \citet{kgw04}. We used the multi-resolution technique to resimulate the galaxy at higher resolution, with a gravitational softening length of 570 pc in the highest resolution region. 
Initially a low resolution dark matter simulation of a comoving 30 h$^{-1}$ Mpc diameter sphere was performed. At z = 0 a region was identified which contained a Milky Way sized halo. We traced the particles that fell into this region including its surrounding regions back to the initial conditions at z = 51.7 and identify the volume that contains those particles. Within this arbitrary volume, we replace the low resolution particles with higher resolution (less massive) particles. The initial density and velocities of the higher resolution particles are computed self-consistently using {\tt Grafic-2} \citep{eb01} taking into account the density fields of a lower-resolution region. We have done 2 levels of this refinement. Finally the simulation of the entire volume is re-run including gas particles (only in the highest-resolution region).
The mass of each gas and dark matter particle is $9.2\times10^{5}$ M$_{\sun}$ and $6.2\times10^{6}$ M$_{\sun}$\footnote{Note that although our simulation was first published in \citet{bkg05} the resolution of this simulation is comparable to the recent Aquila comparison project \citep{swp12}. The simulation also includes an equally or more sophisticated chemical evolution model when compared with recent studies.}. 
The properties of our simulated galaxy is summarized in Table 1.
The second column represents the virial mass; the third column represents the virial radius; fourth and fifth columns represent the mass of each gas and dark matter particle in the highest resolution region and sixth and seventh columns are the softening lengths in that region. The cosmological parameters for the simulation are presented in Columns 8$-$10.  $\Omega_0$ is the total matter density fraction, $h_0$ is the Hubble constant (100 ${\rm kms^{-1}Mpc^{-1}}$) and $\Omega_{\rm b}$ is the baryon density fraction in the universe.
The age of the universe is 13.5 Gyr in our simulation.

\section{Results}
\label{res-sec}

We set our disk sample to live in the cylindrical range 7.0 $<$ R $<$ 10.0 kpc. There are two primary reasons for this restriction. Firstly this is a similar range to that used in recent observational studies. Secondly, this range should minimize the contamination from the bulge of the galaxy. 
In the vertical direction, we limit the sample to 0.0 $<$ $|z|$ $<$ 3.0 kpc for determining radial gradients.
For vertical gradients, we fit in the range 1.0 $<$ $|z|$ $<$ 3.0 kpc. Again, this is mainly to allow for qualitative comparison with previous observational papers \citep[e.g.][]{ccz12}. 
We separate our stellar sample by their positions (in bin sizes of 1 kpc) in order to disentangle the effect of one gradient on the other.
By taking the median points in each bin, and then determining the gradients from these median values, we reduce the effect of 
any outliers in our sample.  
Note that in this paper, 
we have restricted our sample to include only positively rotating insitu stars, which we define to be born within a radius of 20 kpc from the center of our galaxy and to be located within our chosen region at the final timestep \citep[see][for more details]{rah11}. 
By doing this, we eliminate the halo and bulge component and focus only on the disk stars formed in-situ for our comparison with the MW observations.

\begin{table}
 \centering
\begin{minipage}{70mm}
\caption{Radial (vertical) [Fe/H] and [O/Fe] gradients for various heights (radial distances from the galactic center). Note: [O/Fe] is being used as a proxy for [$\alpha$/Fe].}
\begin{tabular}{lllll}
\hline
\multicolumn{5}{c}{Radial abundance gradients (dex ${\rm kpc}^{-1}$)} \\
\hline
\hline
$|z|$ Height & $N$ & Age & Gradient & Gradient \\ 
(kpc) & Stars & (Gyrs) & [Fe/H] & [O/Fe] \\ \hline
 0.0$-$1.0 & $1599$ & 3.42 & $-0.058$ & $+0.001$ \\ \hline
1.0$-$2.0 & $802$ & 4.21 & $-0.003$ & $-0.015$ \\ \hline
2.0$-$3.0 & $394$ & 5.56 & $+0.022$ & $-0.026$ \\ \hline
\multicolumn{5}{c}{Vertical abundance gradients (dex ${\rm kpc}^{-1}$)} \\ \hline
\hline
R Distance & $N$ & Age & Gradient & Gradient \\ 
(kpc) & Stars & (Gyrs) & [Fe/H] & [O/Fe] \\ \hline
 7.0$-$8.0 & $522$ & 4.06 & $-0.156$ & $+0.029$ \\ \hline
8.0$-$9.0 & $382$ & 4.35 & $-0.171$ & $+0.049$ \\ \hline
9.0$-$10.0 & $292$ & 4.15 & $-0.105$ & $+0.006$ \\ \hline
\end{tabular}
\end{minipage}
\end{table}

\begin{figure}
\centering
\includegraphics[angle=270,width=\hsize]{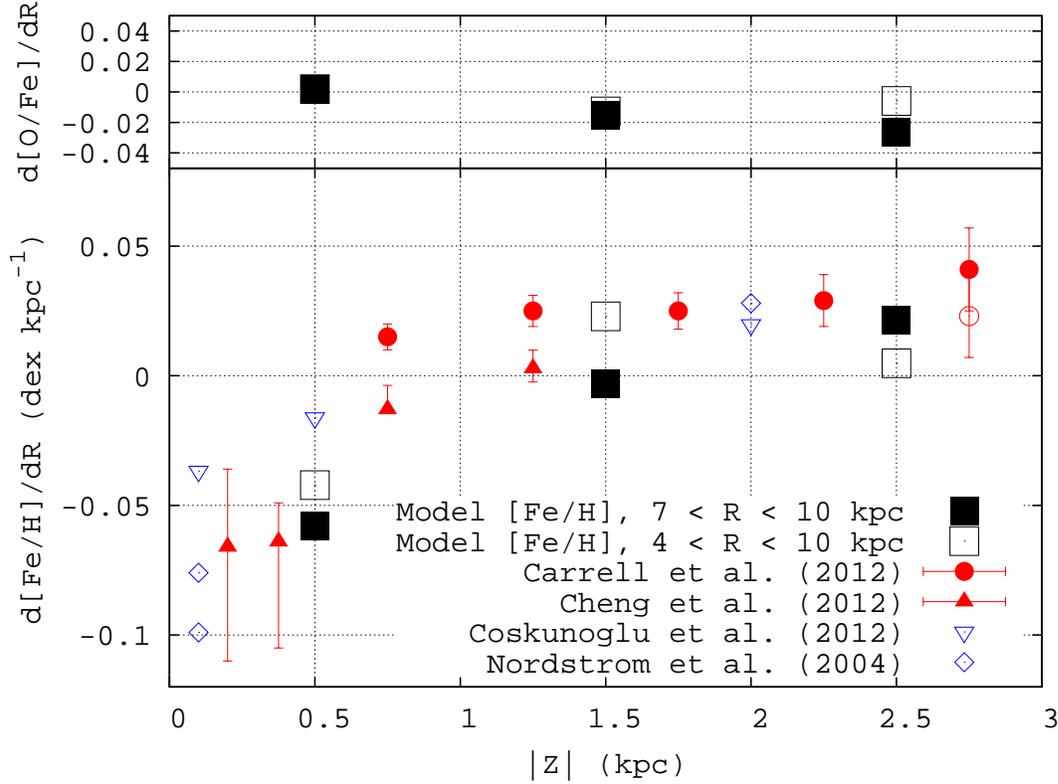}
\caption{
Radial [Fe/H] and [O/Fe] gradients for different heights above the galactic plane.}
\label{totsfr-fig}
\end{figure}

\subsection{Radial Gradients}
\label{comp-acc-ins}

The resulting radial metallicity gradients are shown in Table 2 and Fig.~\ref{totsfr-fig} shows the radial metallicity gradient as a function of average $|z|$.
The radial metallicity gradients evaluated in a larger radial range (4.0 $<$ R $<$ 10.0 kpc) are overplotted for a comparison. We find that the results follow the same general trends in the two different radial bins, except at the most distant vertical height bin. 
Overplotted in red with error bars are the recent results from \citet{che12} and \citet {ccz12} for stars in the Galactic disk.
Other observations \citep{nma04,cab12} are shown in blue. Whilst both of these are solar neighborhood values, they selected $\it{thick-disk-like}$ stars using either kinematics or ages and so we have placed these points for their different selection criteria at roughly where they would belong. This is why they do not have error bars.
The results of these previous publications from the MW are shown to help provide a qualitative comparison.
We stress that it is not our aim to make a quantitative comparison since our simulated galaxy is different from the MW \citep[cf.][]{rah11} and so we should not expect the two to have the same absolute values for the gradients. We can however compare the general trends and make suggestions as    to the possible causes of these trends in our Galaxy.

We find a clear negative radial metallicity gradient close to the plane. As we move up in $|z|$, the gradients become shallower (less and less negative). The final vertical height bin (2.0 $<$ $|z|$ $<$ 3.0 kpc) has a (small) positive gradient. Interestingly the observations of radial metallicity gradients in the MW at different vertical heights suggest a similar trend. 

The top panel of Fig.~\ref{totsfr-fig} shows the radial [$\alpha$/Fe] gradient as a function of average $|z|$, where we have used O as a proxy for the $\alpha$-elements. Note we could also have used Mg or other $\alpha$-elements and the results would have qualitatively been the same.
The radial [$\alpha$/Fe] gradients follow the reverse pattern to [Fe/H]. Initially, they are slightly positive and become negative moving up in $|z|$. This is not surprising since as the Fe abundance increases, the [$\alpha$/Fe] abundance is expected to decrease. 
It will be interesting to see whether current and future observations from the MW and other disk galaxies follow similar trends in radial [$\alpha$/Fe] gradients as a function of $|z|$.

\begin{figure}
\centering
\includegraphics[angle=270,width=\hsize]{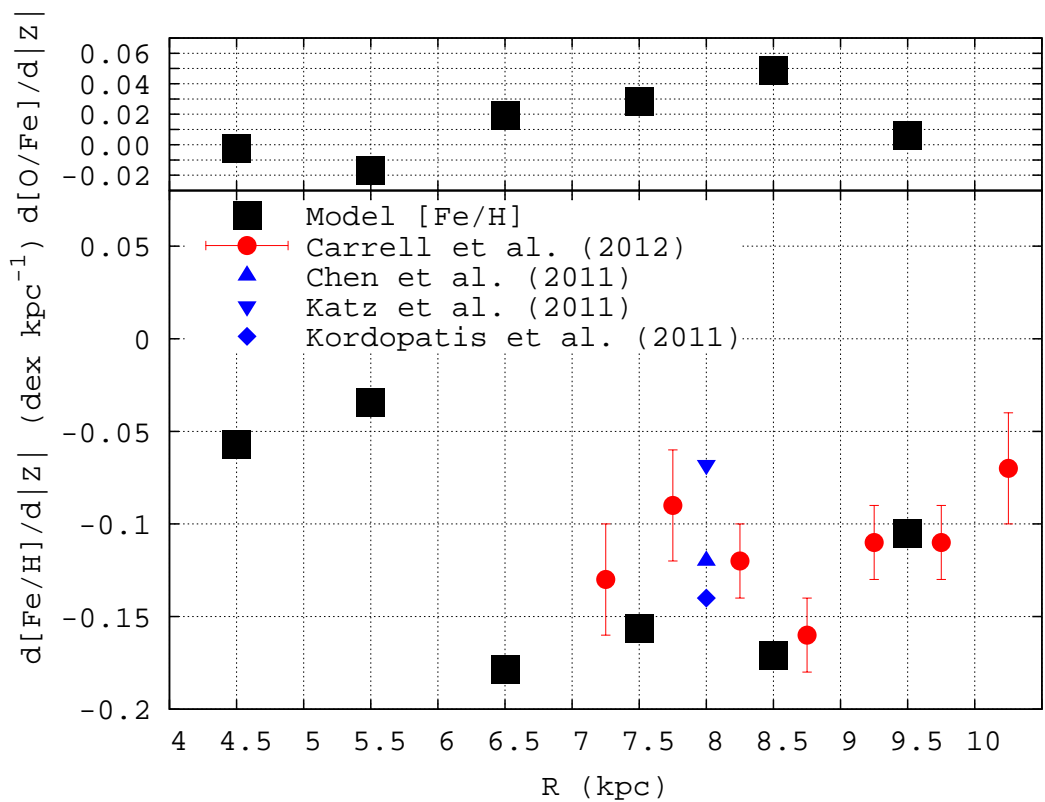}
\caption{
Vertical [Fe/H] and [O/Fe] gradients for different radial distances from the galactic center.
}
\label{luminmap-fig}
\end{figure}

\subsection{Vertical Gradients}
\label{VG}

We also examined the vertical metallicity gradients in different radial bins ranging from 4.0 $<$ R $<$ 10.0 kpc (comparing with observations from 7.0 $<$ R $<$ 10.0 kpc) from the galactic center, fitting in the range 1.0 $<$ $|z|$ $<$ 3.0 kpc in a similar fashion to what was done for the radial gradients. Table 2 shows that there are fewer stars in each bin when we bin in radial distance to fit for vertical gradients. 
Fig.~\ref{luminmap-fig} shows the vertical metallicity gradients as a function of radial distance from the galactic center. Overplotted with error bars are the recent observations of \citet{ccz12} from the MW disk. Other recent observations \citep{ksc11,krd11,czc11} are shown in blue and without error bars since they do not have distances for their values (or are from local samples). We have placed these observations at the solar radius, which is assumed to be 8 kpc.
We find negative vertical metallicity gradients at all radii.
The metallicity gradient at the largest radial distance
is less negative compared to the adjacent inner regions. 
The top panel of Fig.~\ref{luminmap-fig} shows the vertical [$\alpha$/Fe] gradient as a function of radius from the galactic center. Again we see that the [$\alpha$/Fe] gradients follow the reverse pattern to [Fe/H].

From Table 2, we see that there is no clear relationship between the mean age of the stars and the radial distance from the galactic center, for the calculation of vertical gradients. The age increases in the first two radial bins. The most distant radial bin contains younger stars than the middle bin. However, these differences are marginal. 
We remind the reader that the vertical gradients (and hence mean ages) were calculated at heights above 1 kpc.

\section{Discussion: origin of the positive radial metallicity gradient}
\label{age-analysis}

The metallicity distribution, especially in the Galactic thick disk is still an unresolved issue.
Hence, in recent years, with improving observational capabilities, there has been an increasing number of measurements of metallicity gradients in the Galactic disk. \citet{nma04} found that stars likely belonging to the thick disk had a positive gradient in the radial direction. \citet{apb06} found the gradient to be flatter and consistent with 0 dex kpc$^{-1}$. More recent determinations using various methods have also suggested a positive radial gradient \citep[e.g.][]{ksc11,krd11,cab12,ccz12}. 
There is less certainty for vertical metallicity gradients with, on the one hand, \citet{apb06} finding no gradient and, on the other, \citet{czc11} finding a strongly negative gradient. Most recently, \citet{ccz12} used a large sample of F, G, and K dwarfs selected from SDSS Data Release 8 to analyze how the vertical metallicity distribution in the thick disk varied with radial distance from the Galactic center, finding negative gradients but no obvious trends as a function of radial distance.  
The variation in the observed gradients are due to differences in both the samples and methods.
Some \citep[e.g.][]{nma04} are local samples, which have good distance and abundance determinations but a limited range. For the results from common samples (SDSS) there are differences in how both the distance and abundance are determined (and there is an effect of one on the other in addition). Also, some use a guiding radius (determined from orbit tracing, which requires the full space motion) and some just use the current radius. Further observations of larger and more uniform samples of stars are required to make this important observational signature more clear.

\begin{figure}
\centering
\includegraphics[angle=0,width=\hsize]{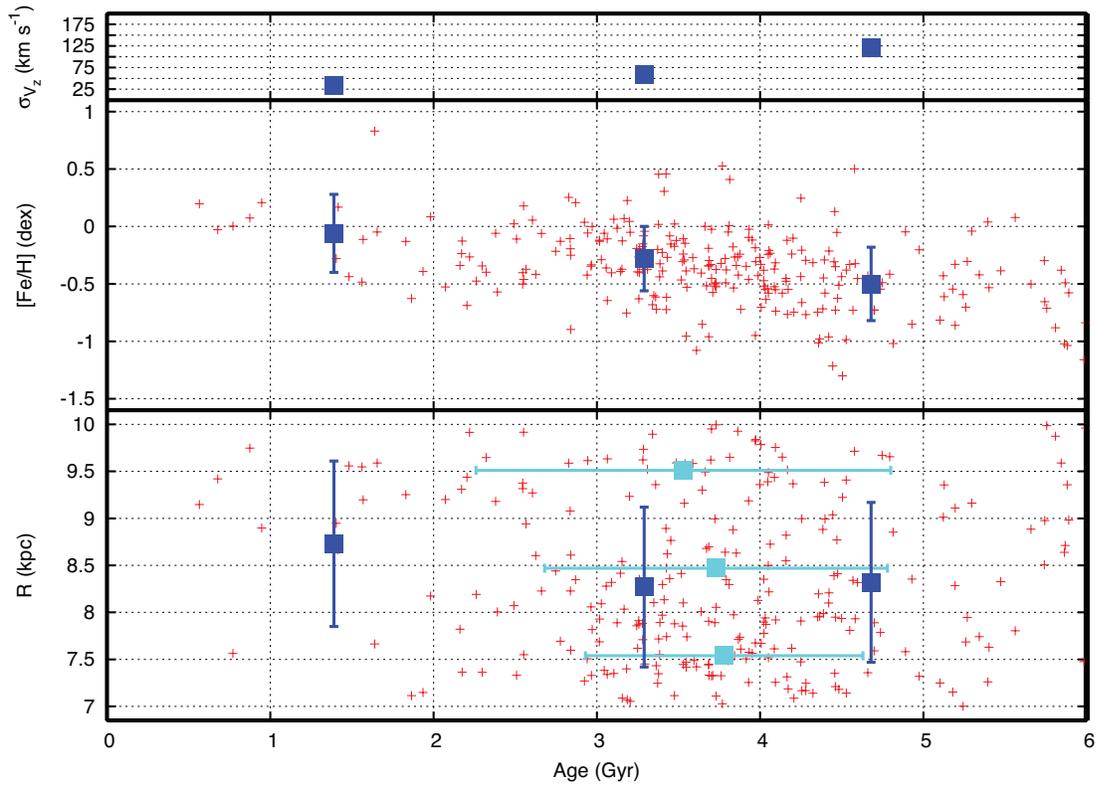}
\caption{
The relationship between age, radial distance, [Fe/H] and velocity dispersion for disk stars located between 2 $<$ $|z|$ $<$ 3 kpc. Note we do not include stars older than 6 Gyr since our simulated disk only formed within the past 6 Gyr. 
The large blue squares show the mean binned in age.
For the bottom panel, as well as binning in the x-direction (age), we also binned in the y-direction (i.e. radius) to get the mean (shown as cyan squares). We see that the youngest disk stars are found at larger R.
}
\label{fehvst-fig}
\end{figure}

From Fig.~\ref{totsfr-fig} we show that the 2 $<$ $|z|$ $<$ 3 kpc bin has a positive radial metallicity gradient in qualitative agreement with recent observations from the MW. Therefore, it is interesting to explore the reason why our simulated disk has a positive radial metallicity gradient at this high vertical height.
Fig.~\ref{fehvst-fig} shows the relationship between the age, metallicity, radial distance, and vertical velocity dispersion, $\sigma_z$, for our disk stars in the 2 $<$ $|z|$ $<$ 3 kpc bin. Note that we have only shown the stars younger than 6 Gyr since this is when the majority of disk stars began to form in our simulated galaxy \citep[see][]{rah11}.
We also analyzed stars older than 6 Gyr. This group was a small minority of the disk population and was found to contain no trends.
Fig.~\ref{fehvst-fig} shows that the younger stars in this sample shows higher metallicity, i.e. age-metallicity relation, and lower velocity dispersion, i.e. age-$\sigma_z$ relation. 

\begin{figure}
\centering
\includegraphics[angle=0,width=\hsize]{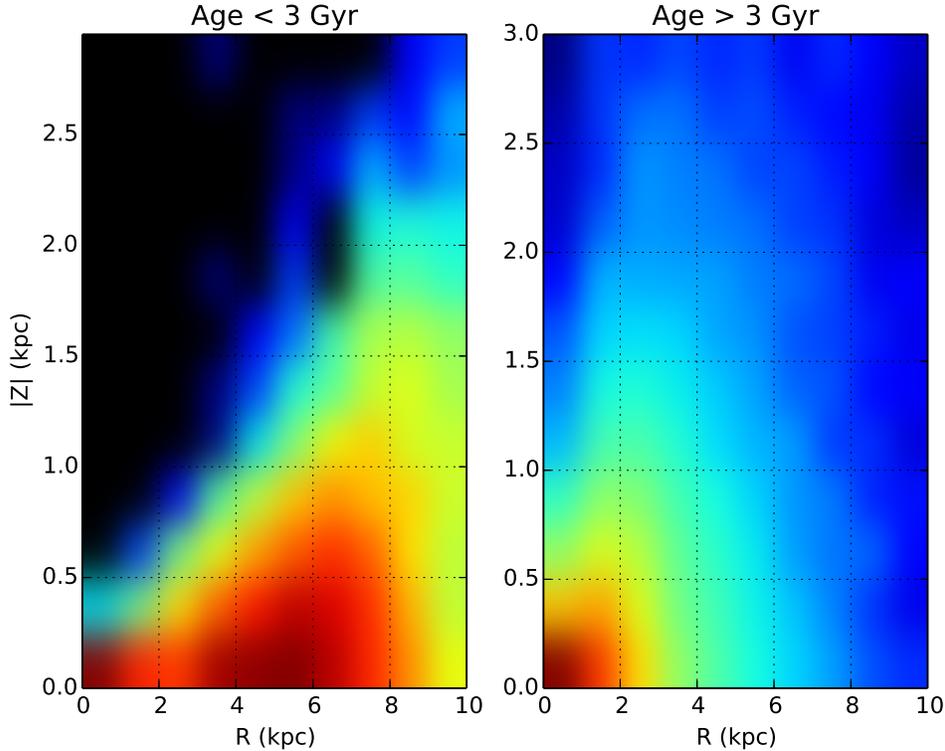}
\caption{
The distribution of positively rotating (disk) insitu stars in the $|z|$ vs. R plane shown as a smoothed density plot, split between young and older stars. Bright (red) is high and dark (black) is low density. From the left panel, we see that there is a hint of flaring at the outermost radii.
}
\label{densitymap-fig}
\end{figure}

Since the star particles younger than 3 Gyr display higher metallicity in Fig.~\ref{fehvst-fig}, we have divided the sample into two groups, 
younger (age $<$ 3 Gyr) and older (age $>$ 3 Gyr) star particles. Fig.~\ref{densitymap-fig} shows the density distribution of the younger and older stars. Because the older stars have high velocity dispersion, they can reach $|z|>2$ kpc at any radius. On the other hand, because the younger star particles have lower velocity dispersion, and the restoring force due to the gravitational potential of the disk is high at the inner radii, the younger stars are confined within lower vertical heights ($|z|<1.5$ kpc) in the inner region. However, because at the outer radii the restoring force from the disk is weaker, the younger stars can reach higher vertical heights. Fig.~\ref{densitymap-fig} shows a `flaring' feature \citep[e.g.][]{bkw12,bkw13,rds12} for the younger stars. In our simulation, the relative excess of younger, lower $\sigma _{z}$ and higher [Fe/H] stars found at larger R as a result of this flaring is the cause of the positive radial metallicity gradient at the high vertical height.

\citet{hdl13} demonstrate that the MW thick disk is old, but has a clear age-metallicity relation, with younger thick disk stars having higher [Fe/H] (and lower [$\alpha$/Fe]). \citet{hdl13} also showed that the younger thick disk stars have lower velocity dispersion, $\sigma _{z}$. We can consider that the MW thick disk stars have a similar chemodynamical origin to our sample of stars in Fig. 3, despite their age difference. Then, we expect that because the older thick disk stars in the MW have higher velocity dispersion, they can reach high $|z|$ at all radii. On the other hand, more younger thick disk stars that have a smaller $\sigma _{z}$ can reach high $|z|$ only at the outer radii, because the gravitational potential due to the stellar disk is smaller in the outer region, i.e. flaring. As seen in our simulation, this can explain the positive radial metallicity gradient observed in the MW thick disk at high vertical heights. 

There are several factors that could potentially affect our results which we have tried to account for and minimize.
Firstly, our simulated galaxy suffers from classical overcooling problems.
At later times when much of the gas has cooled out of the disk, the disk can easily be ``over-enriched" by feedback events. This means that we should not rely on the absolute values of our abundances in the disk. Hence, we focus only on the abundance {\it trends} along the disk and draw conclusions from these trends, which should remain robust. 
The overcooling issue results in spheroid over-production of stars.
This is significant since as we move up out of the plane, our sample becomes increasingly prone to spheroid contamination. 
We minimize any possible contamination by setting our disk sample to live in the cylindrical range 7.0 $<$ R $<$ 10.0 kpc, and only including the positively rotating insitu stars.
We also note that we expect a negative metallicity gradient for such a spherical stellar component \citep[e.g.][]{tsb13} and so our key results are unlikely to be due to contamination.

Resolution is another important issue which can affect the results. Specifically, the vertical gradients were determined using quite low number statistics. This makes them less reliable than the radial gradients (which are the focus of this paper). Also, at larger vertical heights above the plane and radial distances from the galaxy center, our sample becomes less reliable due to the lower number statistics. 
Obviously going to higher resolution will be beneficial, however we have tested our results using different bin sizes and reach similar conclusions.

In this paper, we have demonstrated a possible scenario for the observed positive radial metallicity gradient in the MW thick disk. 
An obvious limitation of this study is that we have used only one chemodynamically-simulated galaxy.
One may ask the question, how different can these gradients be? Are they unique or expected for all MW-like galaxies? To address these questions, we are in the process of carrying out a larger and more detailed comparison project where we will use several cosmological disk galaxies simulated with different codes and under different cosmologies, which should provide us with important constraints and boundary conditions for models of Galactic formation.

\section*{Acknowledgments}

AR and KC acknowledge the support of the National Astronomical Observatories and Chinese Academy of Sciences. AR is supported by the Chinese Academy of Sciences Fellowships for Young International Scientists and by the National Natural Science Foundation of China (NSFC) under Grant Number: 11333003. KC is supported by the NSFC under grant Nos. 11233004 and 11243004.
AR acknowledges the guidance and helpful support of Brad Gibson, Chris Brook and Shude Mao.
We acknowledge CfCA/NAOJ and JSS/JAXA where the numerical computations for this paper were performed.


\end{document}